\documentclass[12pt,preprint]{aastex}
\usepackage[]{times, graphicx}
\usepackage[pdftex]{color}
\usepackage{hyperref}
\received{December~12, 2008}
\revised{March~14, 2009}
\accepted{April~2, 2009}

\newcommand{\alfven}{Alfv\'{e}n}

\newcommand{\pref}{\protect\ref}
\newcommand{\oratio}{\mbox{O$^{7+}$/O$^{6+}$}}

\begin{document}

\shorttitle{Source of the Solar Wind}
\shortauthors{Leamon \& McIntosh}
\title{How The Solar Wind Ties To Its Photospheric Origins}


\author{Robert J. Leamon\altaffilmark{1}, Scott W. McIntosh\altaffilmark{2}}
\email{robert.j.leamon@nasa.gov, mscott@ucar.edu}
\altaffiltext{1}{Adnet Systems Inc., NASA Goddard Space Flight Center, Code 671.1, Greenbelt, MD 20771 USA}
\altaffiltext{2}{High Altitude Observatory, National Center for Atmospheric Research, P.O. Box 3000, Boulder, CO 80307}

\begin{abstract}
We present a new method of visualizing the solar photospheric magnetic field based on the ``Magnetic Range of Influence'' (MRoI). The MRoI is a simple realization of the magnetic environment in the photosphere, reflecting the distance required to balance the integrated magnetic field contained in any magnetogram pixel. It provides a new perspective on where sub-terrestrial field lines in a Potential Field Source Surface (PFSS) model connect to the photosphere, and thus the source of Earth-directed solar wind (within the limitations of PFSS models), something that is not usually obvious from a regular synoptic magnetogram. In each of three sample solar rotations, at different phases of the solar cycle, the PFSS footpoint either jumps between isolated areas of high MRoI or moves slowly within one such area.
Footpoint motions are consistent with Fisk's interchange reconnection model.

\begin{center}
Received December 12, 2008; accepted April 2, 2009.
\end{center}

\end{abstract}


\section{Introduction}

That the speed of earth-directed solar wind and the region from which it originates is tied to the large scale configuration of the photospheric magnetic field has been understood for at least 40 years. \cite{SchattenEA69} and \cite{AltschulerNewkirk69} originated the concept of the Potential Free Source Surface (PFSS) model. PFSS models assume that the coronal magnetic field is quasi-stationary and can, therefore, be described as a series expansion of spherical harmonics. PFSS models have become the staple of solar wind prediction models that have increased in complexity over the years. For example, \cite{WangSheeley90} found an empirical inverse correlation between the super-radial expansion factor of a magnetic flux tube between the photosphere and source surface with the resulting solar wind speed observed at 1~AU. \cite{WangSheeley91} then demonstrated that the observed correlation was consistent with simple wind acceleration models involving \alfven\ waves 
\cite[e.g.,][]{LeerEA82}. 
\cite{ArgePizzo00} made further enhancements by accounting for stream-stream interactions of the wind en route to 1~AU. The end result was a fairly accurate predictive model, that runs stably and continuously, for near real-time space weather 
forecasting.\footnote{\href{http://www.sec.noaa.gov/ws/}{{\tt http://www.sec.noaa.gov/ws/}}}
Similar PFSS-based predictive tools have been developed recently by applying these techniques \cite[]{SchrijverDeRosa03} to observations from SOHO/MDI 
\cite[]{ScherrerEA95_short}.\footnote{\href{http://www.lmsal.com/forecast/}{\tt http://www.lmsal.com/forecast/}}
We do note, however, that PFSS models make the assumption that the corona is current-free 
(enabling the use of spherical harmonics), a flawed assumption in a significant fraction of the quiet corona that becomes worse in active regions.


The ``Magnetic Range of Influence" (MRoI) was conceived by \cite{McIntoshEA06}
as a diagnostic to understand the partitioning of the \ion{Ne}{8} Doppler velocities observed by SOHO/SUMER \citep[][]{WilhelmEA95_short} in a large equatorial coronal hole \cite[see, e.g., Fig.~3 of][]{McIntoshEA06}. 
The MRoI is simple realization of the magnetic environment in the photosphere, reflecting the distance required to balance the integrated magnetic field contained in any pixel in the magnetogram. 
In practice, it is calculated by repeated convolution of the input magnetogram with a circular kernel of increasing radius.
While the MRoI contains no directional information, it allows the partitioning of the magnetic field into open and closed regions. 
When the MRoI is large, the magnetic field at that point is largely unbalanced and the magnetic environment is effectively ``open." 
\cite{McIntoshEA06} noticed that, in a coronal hole where the MRoI is large, the field imbalanced and open significant outflow is seen in \ion{Ne}{8}, while, in the quiet sun the MRoI is typically small, balanced and closed, the \ion{Ne}{8} mean Doppler velocity is close to zero and the line intensities are a factor of~3 higher than in the open regions. 
Put another way, magnetic environment leads to a preferential energy balance in the upper transition region plasma where the magnetically closed regions are dominated by plasma heating while open regions are dominated by kinetic energy showing decreased emission and strong outflow. 
This is supported by another perennial signature of coronal hole outflow, reduced hot oxygen charge states \cite[the ratio \oratio{};][]{OgilvieVogt80}.
 
In this Letter, we will show that high MRoI is not just indicative of outflow on small (supergranular) scales but also on the largest scales. By computing the MRoI for synoptic magnetograms, we can show where the footpoint of earth-directed solar wind from a PFSS model attaches. This approach will increase our understanding of the footpoint's movement, its basal energetic state and will result in the improved interpretation of in situ wind measurements.

\section{Observations}

%
%



\notetoeditor{Per the aastex style guide's request, we have presented the synoptic map and in situ time series panels of each figure as separate eps files.  In the final Journal typesetting, we would like them aligned and the same size (width).  We can do this ourselves in Illustrator, or leave it to the Journal's graphics experts\ldots}

\subsection{July--August, 2003: CR~2005}

\begin{figure}
\epsscale{0.7}
\plotone{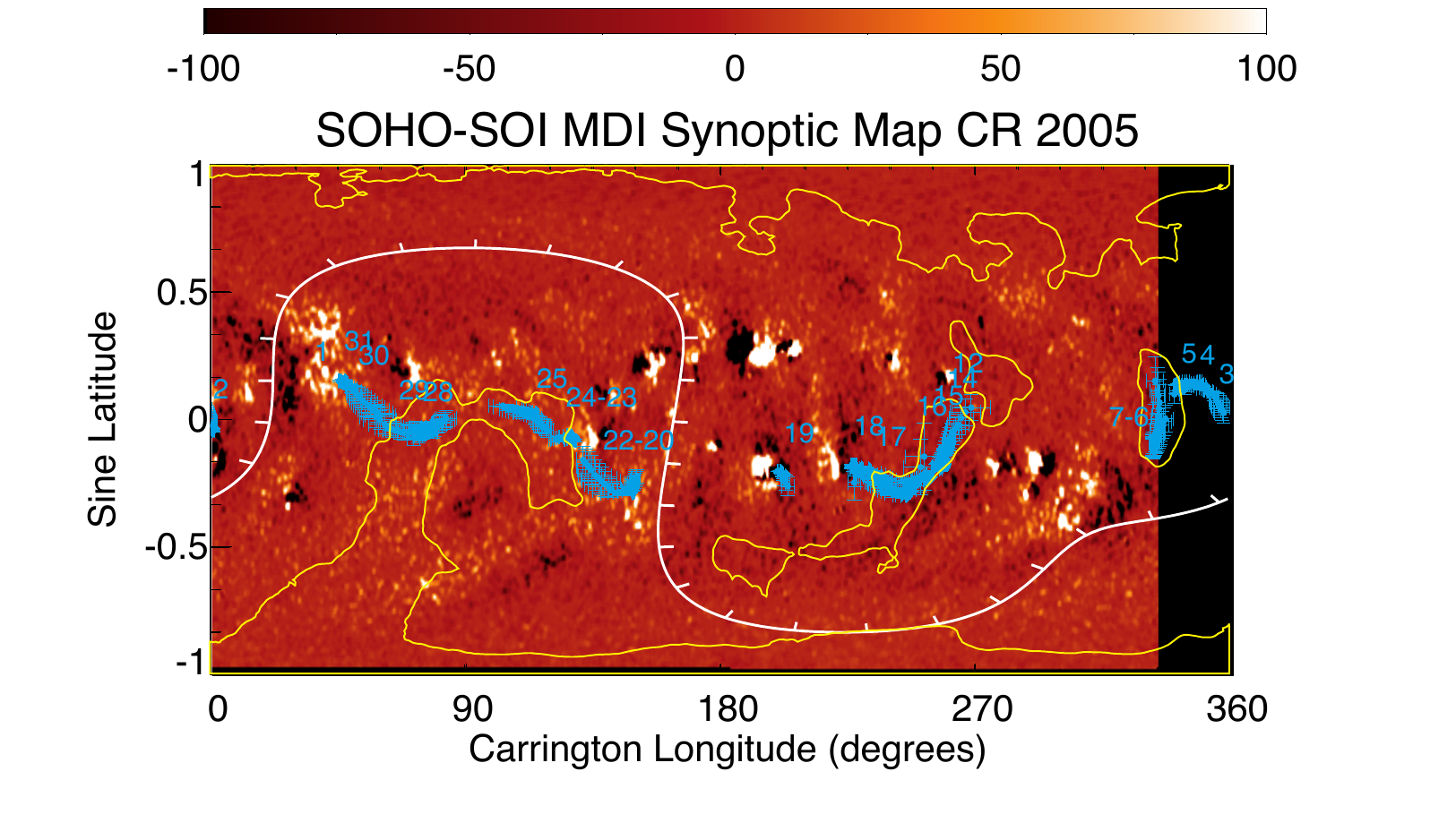}
\epsscale{0.65}
\plotone{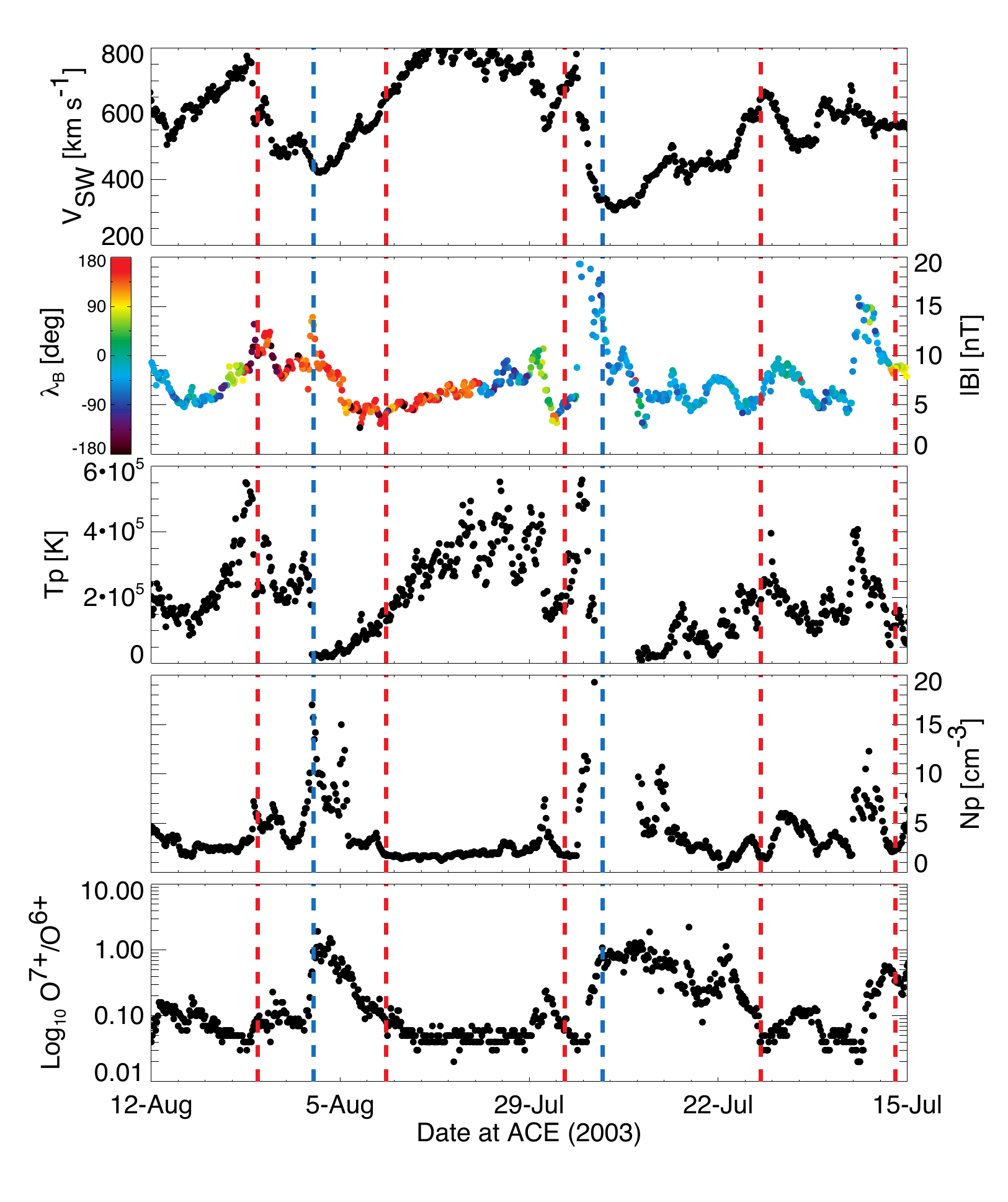}
\caption{Upper panel: Synoptic photospheric magnetic field map for CR~2005.  
               See text for annotations.
               Lower panels: Time series of solar wind speed, magnetic field strength and azimuth, 
               proton temperature and density, and oxygen charge state ratio \oratio{}
               observed in situ by ACE\@.
               Note that time runs from right to left in these panels, as in the synoptic magnetogram.
               \label{fig:2005is}
               }
\end{figure}

This is the interval studied in our forecast paper \cite[]{LeamonMcIntosh07}, which more-or-less corresponds to Carrington Rotation~2005, contains two equatorial coronal holes (ECHs) on opposite sides of the sun that have opposite magnetic polarities. The upper panel of Fig.~\ref{fig:2005is} shows a synoptic magnetogram for CR~2005 downloaded from the MDI archive at Stanford 
University\footnote{\href{http://sun.stanford.edu/synop/}{{\tt http://sun.stanford.edu/synop/}}} and downsampled by a factor of four - from $3600 \times 1080$~pixels to $900 \times 270$~pixels.
The
first 30\degr\ or so of the rotation lacks MDI data. 

The blue dots track the footpoint of the sub-terrestrial field line determined from PFSS extrapolations for the period (one dot per 96~min MDI magnetogram), labeling the progression of time (right to left) at the footpoint closest to noon each day. Repeating the analysis of \cite{LeamonMcIntosh07}, we trace 32 additional field lines (arranged on the perimeter of an ellipse on the source surface with semi-major axes $5\degr \times 2.5\degr$) back to the photosphere. The additional field lines allow us to compute the standard deviation of the separation between their footpoints and that of the sub-terrestrial field 
line---thus providing error estimates for the location of each sub-terrestrial footpoint. The PFSS extrapolation-derived projection of the heliospheric current sheet\footnote{On the disk this is referred to as the heliospheric neutral line, but the two can be used interchangeably.} at the source surface down onto the photospheric magnetogram is shown in white while the boundaries of the coronal holes are shown in yellow. 
The coronal hole contours are derived from Kitt Peak spectroheliograms of 
\ion{He}{1} 10830\AA{} \cite[]{HenneyHarvey05}. 

We clearly see that the footpoint  progresses smoothly only in short segments across the map---it then jumps from one segment to the next, with jumps of $\sim 45\degr$ over both crossings of the neutral line. 

The lower panels of Fig.~\ref{fig:2005is} shows in situ data observed at 1~AU by ACE, from top to bottom: the solar wind speed; 
magnetic field strength and (color-coded) azimuth angle $\lambda_B$ \cite[MAG;][]{SmithEA98_ACE}; 
proton temperature \&\ density, and 
Oxygen charge state ratio \oratio{} \cite[SWIMS;][]{GloecklerEA98_short}. 
The time shown in these panels is that {\em at ACE}, and runs from right to left, as in the synoptic magnetogram.
%
The blue dashed vertical lines correspond to the time of the first magnetogram after the 
North-South crossing of the heliospheric current sheet occurred (2003 July 20, 03:10UT)
and the return crossing (2003 August 1, 08:03UT).
{Throughout this Letter we adopt use the convention that blue  lines correspond to crossings of the heliospheric current sheet, while red lines correspond to when the footpoint enters or exits coronal holes.} 
The various red lines correlate well with coronal hole outflow as defined by low \oratio\ ratio and high
$V_{SW}$.
Also, after the shocks (blue dashed lines) show the classic signature of a current-sheet crossing: as the magnitude increases (field-line packing), the field direction rotates smoothly through over $180\degr$.



Again, it is all well and good explaining the correlations between the synoptic magnetogram and in situ data {\em a posteriori\/}, but can we explain why the PFSS footpoint should jump $\sim30\degr$ of longitude into the coronal hole, or the other jumps for that matter?

Herein lies the usefulness of the MRoI: we calculate the MRoI using the synoptic magnetogram that is the first panel of Fig.~\ref{fig:2005is} as input and show the result in Fig.~\ref{fig:2005}. 
We see that the MRoI map is very patchy and that the footpoint jumps from one patch of (relatively) high MRoI to the next. 
The northern and southern activity belts are still immediately apparent, but the two main equatorial coronal holes have the highest values. 
When there is no close ``island'' (within some 30\degr{}) of higher MRoI, the footpoint stays anchored or moves slowly within the island of higher MRoI (e.g., July 12--18), but when there is no real dominant region, the footpoint moves rapidly across the solar photosphere (e.g., July 19--22, where it is also crosses the neutral line).

\subsection{August--September, 1996: CRs~1912--3}


Fig.~\ref{fig:wsm} combines an MRoI map as in Fig.~\ref{fig:2005} with the same interplanetary variables as Fig.~\ref{fig:2005is}, but for the trailing half of CR~1912 and the leading half of CR~1913. This is the original ``Whole Sun Month'' (WSM). 
Since ACE was not launched until August 1997, the in situ data comes from the OMNI database for the plasma and magnetic field data, and the composition data comes from the SMS experiment on WIND \cite[]{GloecklerEA95_short}. 

The lower temporal resolution of the \oratio\ data is apparent, and (perhaps not unrelated), there is less of a clear signature  of coronal hole outflow in the bottom panel of Fig.~\ref{fig:wsm}. 
Again, we show three dashed vertical lines to indicate the reference times of large-scale footpoint motion (from right to left): the footpoint jumps across the heliospheric current sheet from edges of the southern polar coronal hole (PCH) to the northern PCH (1996 August 19, 09:35UT); the footpoint jumps into the narrow extension of the northern PCH---{see, e.g., Fig.~2 of \protect\cite{Zurbuchen07}}---%
colloquially called the 
``Elephant's Trunk''
(1996 August 24, 12:47UT); 
the footpoint leaves the PCH extension, re-crosses the heliospheric current sheet and attaches to the trailing edge of AR~7986 (1996 August 29, 06:23UT).
The changes in solar wind conditions corresponding to the two current sheet crossings and jump into the coronal hole are again clearly visible.

\subsection{March--April, 2008: CR~2068}


The Whole Heliospheric Interval \cite[WHI;][]{GibsonEA08} comes close to the absolute nadir of the solar cycle, when there were weeks with no magnetic regions of any size on the solar disk. It was somewhat of a surprise, therefore, to see the ``train'' of three equally spaced active regions (NOAA ARs~10987, 10988 and 10989) across the disk. Fig.~\ref{fig:2068} again overlays the MRoI with \ion{He}{1} 10830\AA{}-defined coronal hole boundaries, PFSS footpoints and heliospheric neutral line. We see how little magnetic field there is on the disk by the very low MRoI values, and general lack of contrast.

The northern coronal hole is completely unbalanced, as is AR~10987. Indeed, the latter's influence extends from 2008 March 21, when the footpoint jumps by 94\degr{}~(!) of longitude to the lead edge of AR~10987. 
At this time, the Carrington Longitude of the footpoint  (270\degr{}) corresponds to a heliolongitude of W65; the path of the field line through the corona is highly convoluted. 
While the yellow contours suggest  that the footpoint is actually between the core of the active region and the ECH that precedes it, inspection of EUV data 
(and the in situ panels of Fig.~\ref{fig:2068})
suggest that the footpoint is in the ECH.
The footpoint stays close to, or on, AR~10987 until late on 2008 March 29, where the footpoint jumps by 60\degr\ of longitude over the neutral line to the trailing edge of AR~10989.
After 3.75 days connected to AR~10989, the footpoint moves into a large extension of the southern polar coronal hole.
However, the high speed coronal hole outflow catches up to 
the (notably) slow active region outflow, and entry into the coronal hole occurs 
(red dashed line)
just less than two days after the current sheet crossing.  
Both our coronal hole entry and exit predictions are well matched to the observations. 

The agreement of predicted and actual timing of the shocks is less good than the other intervals studied above. The first shock, corresponding to the 90\degr\ jump to AR~10987, is too early by 32.4~hours while the second shock, corresponding to the jump from AR~10987 over the neutral line to AR~10989 is in much better agreement with prediction. There are two obvious explanations for being off by over a day: (1) we fail to adequately account for the (tortuous) path of nascent solar wind through the corona from W65 to the sub-terrestrial point; and (2) the errors introduced from observing W65 with a line-of sight magnetogram leads to errors in the PFSS model---i.e., in reality the footpoint doesn't jump until a day or so later.
Another, related possibility is if the synoptic map used to compute Fig.~\ref{fig:2068} is generated  from thin strips from a large number of magnetograms over the full rotation when each (Carrington) longitude is the central meridian as seen from SOHO.
It fails to account, therefore, for the evolution of AR~10987 in the 5.5~days it takes took to rotate from W65 to disk center. 

\section{Discussion}

In three examples presented here, over all phases of the solar cycle, synoptic MRoI images provide a striking and easy-to-interpret map of the Earth-directed solar wind source region. When there are islands of high MRoI, the footpoint remains connected to that island until another, more ``enticing,'' island rotates closer to the central meridian. We have seen that this battle for magnetic supremacy leads to the solar maximum wind structure that has many staccato jumps due to the distribution of large MRoI regions on the disk, while at solar minimum the wind has a largely repeating structure with the footpoint meandering from one supergranular vertex to another at disk center or in the polar regions. 

We should note that even though the scaling of the MRoI correlates well with nascent outflow velocity in the upper transition region/low corona \cite[i.e., $\simeq 20 \mbox{ km s}^{-1}$  \ion{Ne}{8} Doppler shifts;][]{McIntoshEA06}. We would not expect a quantitative correlation between MRoI and solar wind velocity at 1~AU---both strong active regions and deep coronal holes have large MRoI values, but they give rise to very different winds speeds and composition. However, such verification is beyond the scope of this Letter and is reserved for future work.

Clearly, equatorial coronal holes have a significant impact on the in situ wind parameters observed. Based on the evidence presented in McIntosh et al.\ (2006, 2007) \nocite{McIntoshEA06,McIntoshEA07} the boundary of the equatorial holes observed in the upper transition region is (spatially) abrupt---in that the spectroscopic diagnostics differ dramatically between the open and closed magnetic regions on the scale of a few arcseconds on crossing the boundary. However, what is not clear is the extension of that boundary into interplanetary space, the extension to the inner heliosphere, the effects of interchange reconnection \cite[]{FiskEA99} and the jumping of sub-terrestrial field line footpoint to flux regions with large MRoI. In Fisk's model, the open field line moves over the photosphere a distance that is determined by the size of the interacting loop. One might argue that the steady walk of the footpoint through the coronal holes in all three examples supports Fisk, where any closed loops are likely to be small, but what of the more heterogeneous quiet sun? Can the jumps from one large MRoI region to another (in equatorial holes, active regions or even across the neutral line) be explained by these MRoI regions forming ``basins of attraction'' at significantly larger distances from the Sun than the supergranular scales for which the MRoI was designed?
One might speculate that a ``likelihood of jumping to here'' function would involve the strength of the MRoI at a point, and a diminishing function of distance between the current footpoint and the candidate point, either across the photosphere, or along a loop \cite[i.e., up into the corona, c.f.][]{Fisk03}, 
but much more work is needed (and planned) to fully understand the nature of large-scale footpoint jumps.

\begin{figure}
\epsscale{0.7}
\plotone{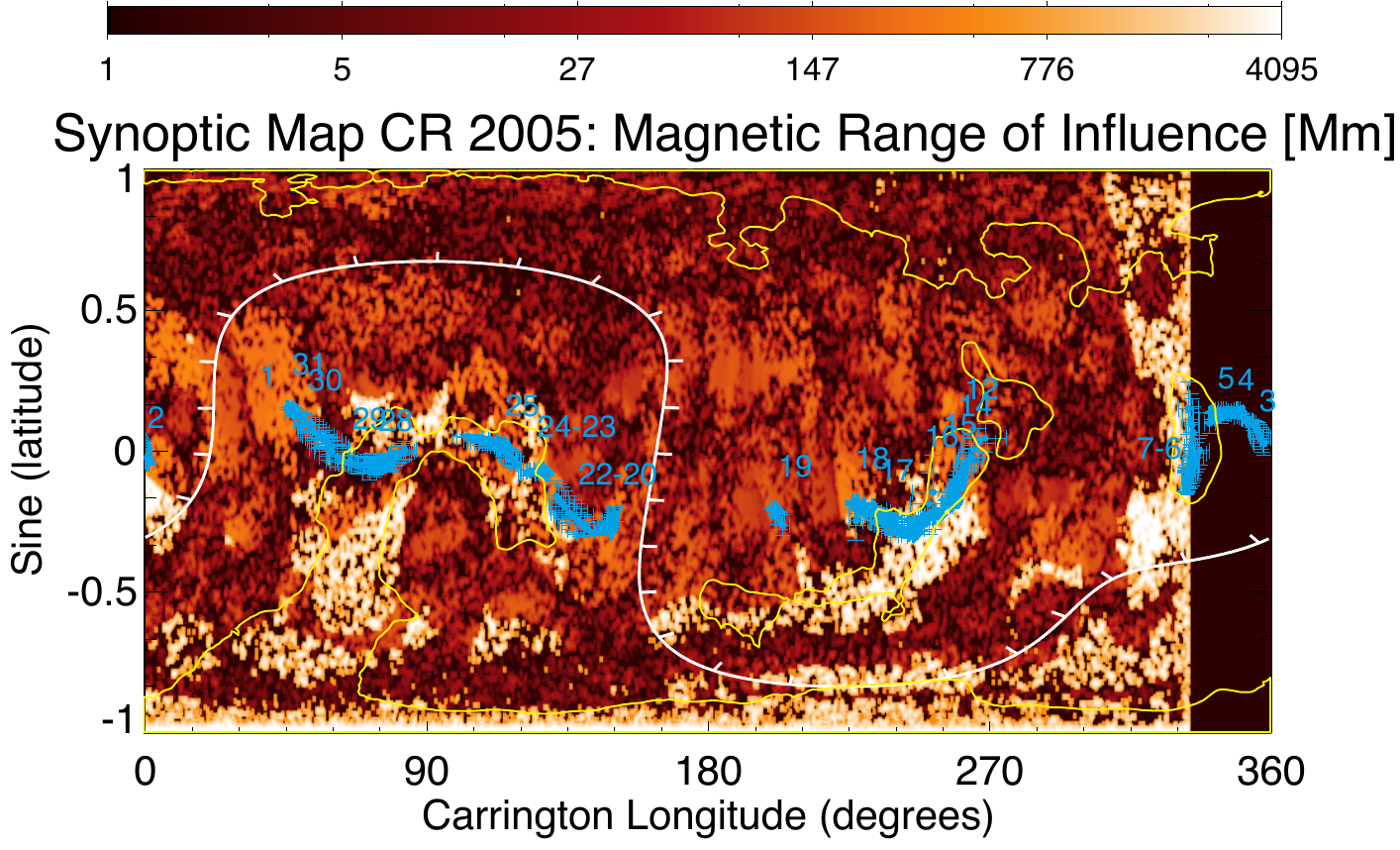}
\caption{The MRoI map from the magnetogram  of Fig.~\pref{fig:2005is}.  
               \label{fig:2005}
               }
\end{figure}

\begin{figure}
\epsscale{0.7}
\plotone{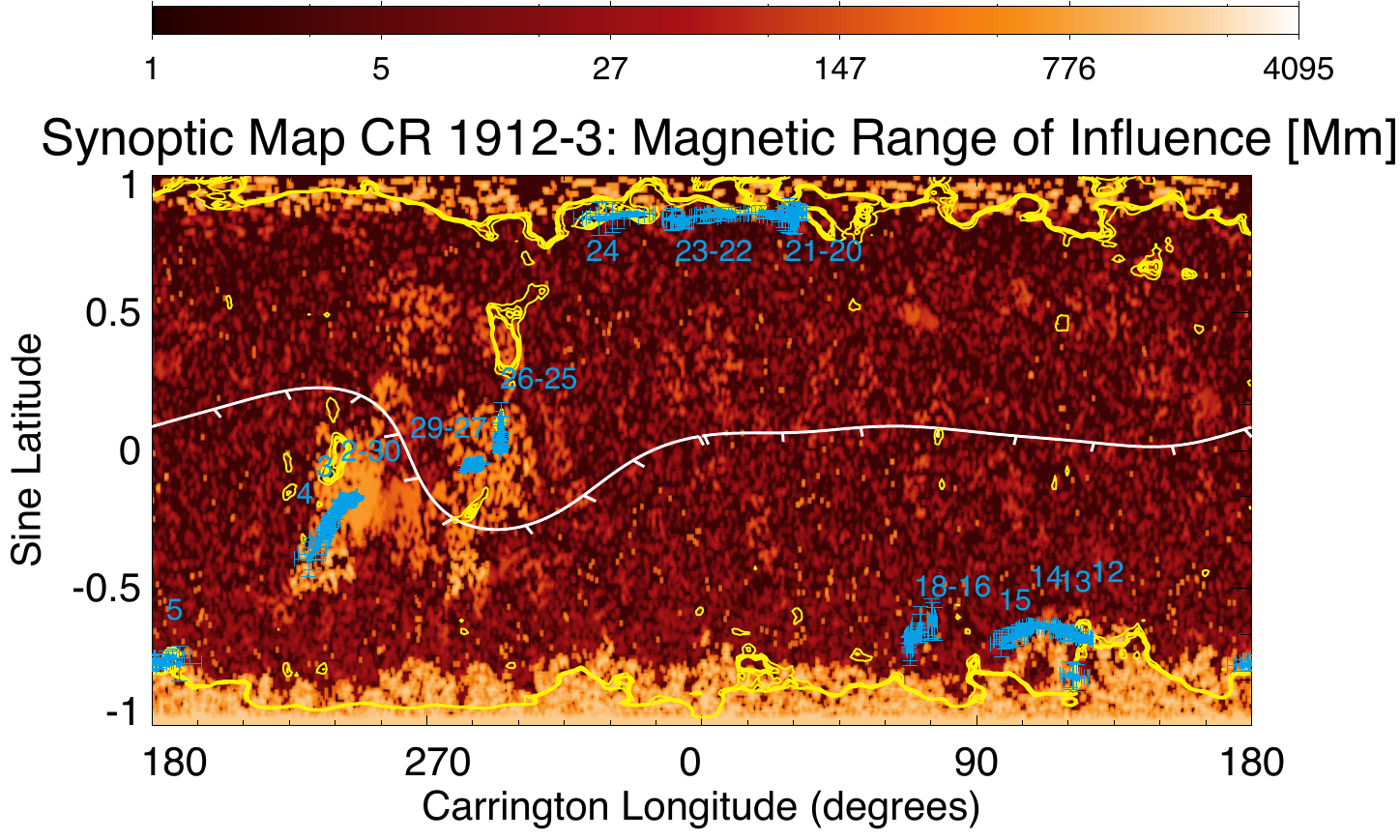}
\epsscale{0.7}
\plotone{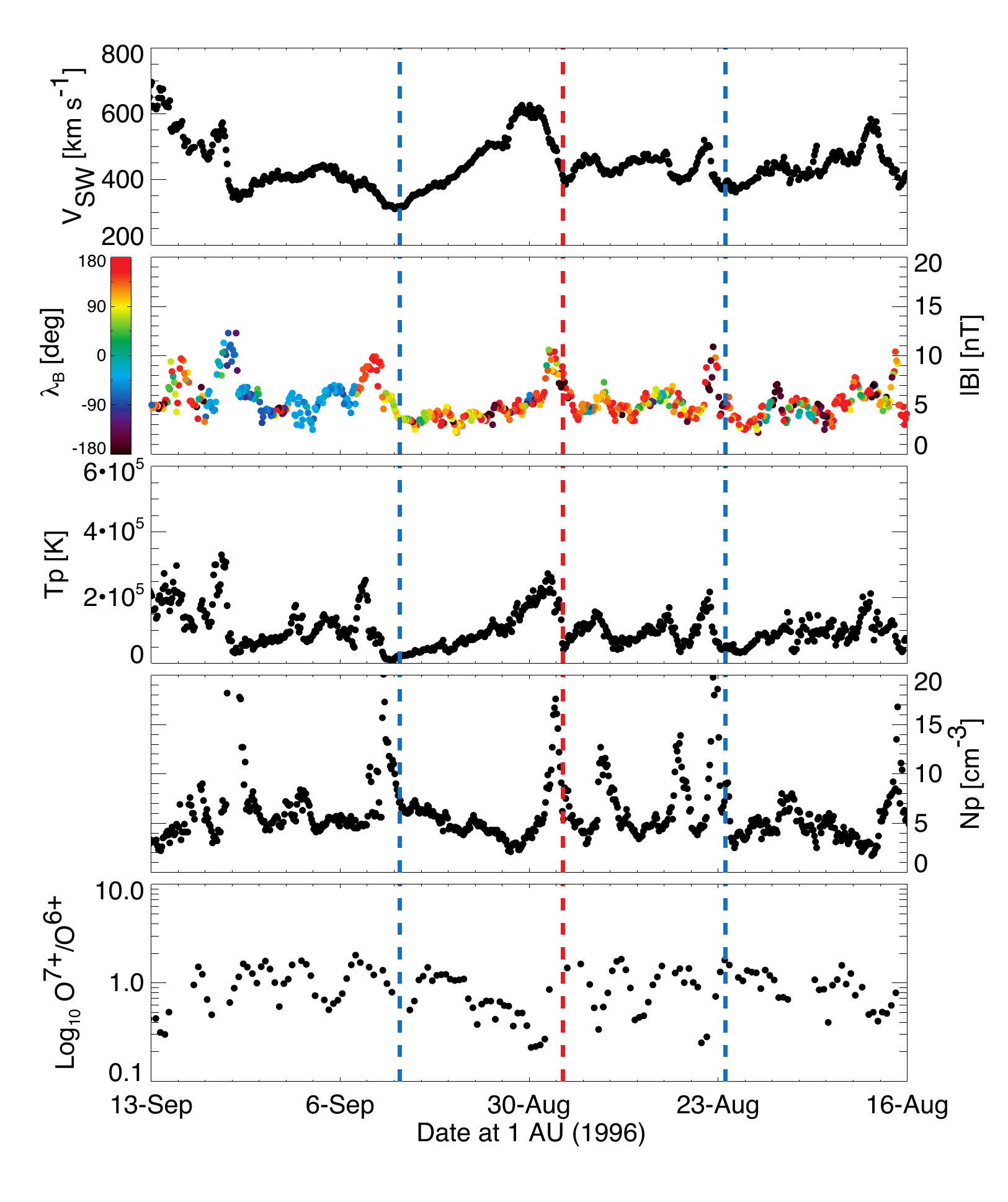}
\caption{As Fig.~\pref{fig:2005is} but for the first ``Whole Sun Month'' interval of August--September, 			1996 and synoptic MRoI map as the upper panel.  Again, time runs from right to left in the 
		in situ data.
 		\label{fig:wsm}
}
\end{figure}

\begin{figure}
\epsscale{0.7}
\plotone{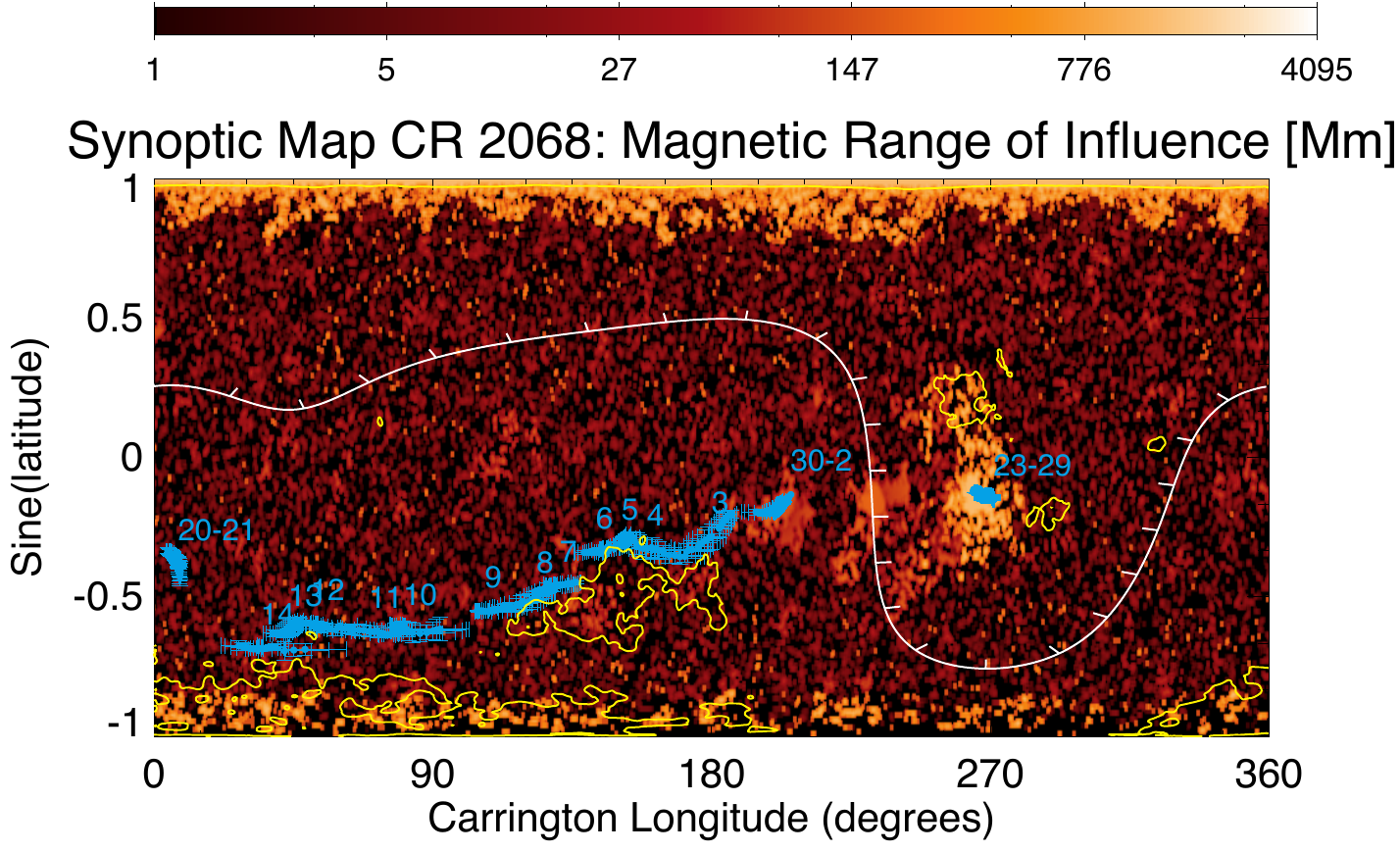}
\epsscale{0.7}
\plotone{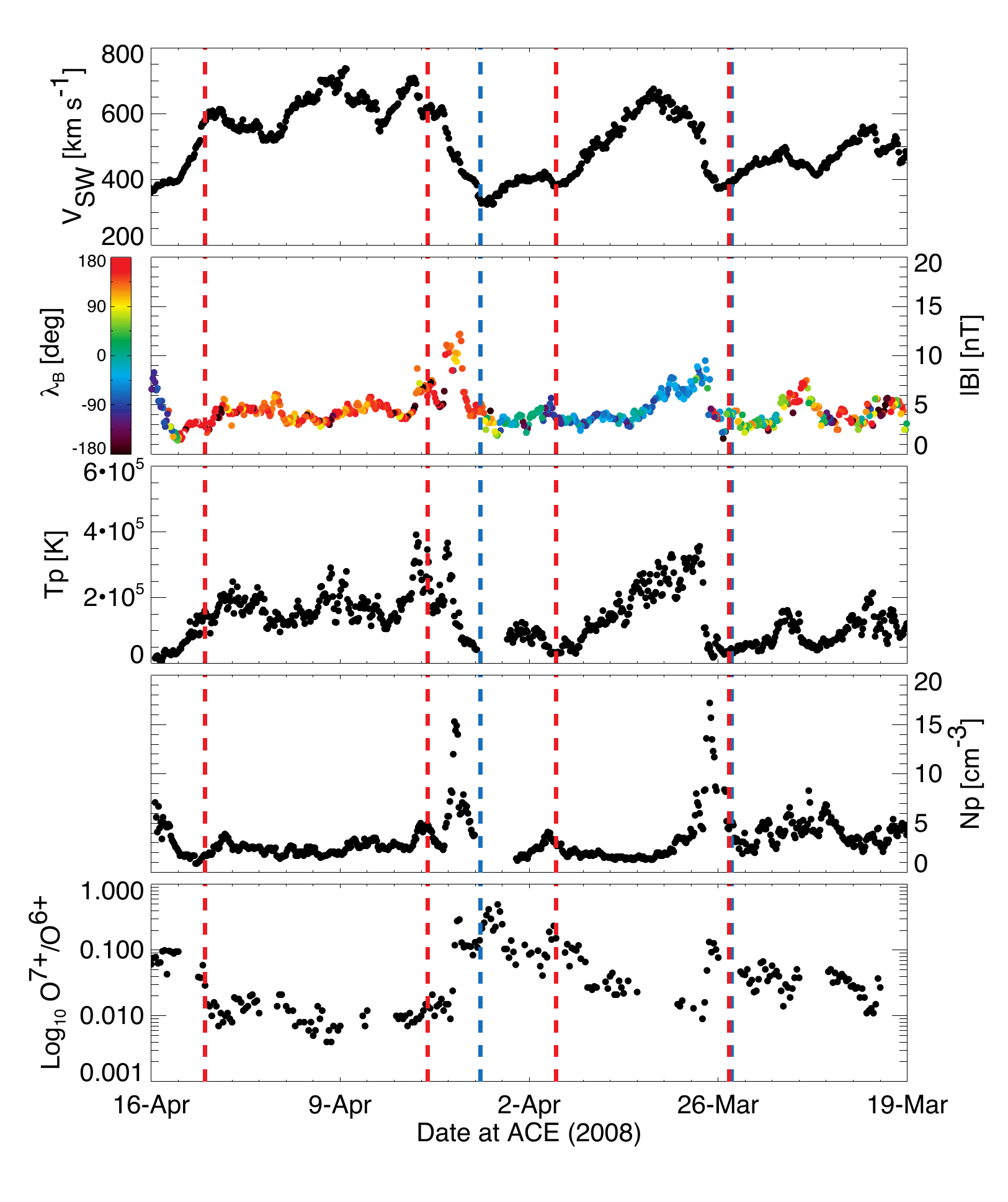}
\caption{As Fig.~\pref{fig:wsm}, but for Carrington Rotation~2068 (the WHI period; March--April~2008).             		\label{fig:2068}
}
\end{figure}

While the biggest drivers of geostorm activity are Coronal Mass Ejections (which no static, synoptic-based model can allow for) we know that the stream-stream interactions caused by the footpoint jumping from point to point at solar minimum can certainly generate shocks with sufficient momentum to impact the geomagnetic indices $D_{st}$,  $K_p$, etc.\ \cite[e.g.,][]{Russell00}. Estimating the timing of these jumps accurately is critical for any predictive model of Space Weather. While the essence of the MRoI jump conditions are not yet known, the MRoI is a visualization tool that offers a great deal of predictive potential; with the study of many epochs we hope to develop predictive intuition and will undoubtedly advance our interpretation and prediction of solar wind conditions observed at 1~AU. As an operational concern, the time needed to generate MRoI maps for automatically updated synoptic magentograms, takes about 20~hours for conditions close to solar max, and about 4~hours for a solar minimum magnetogram on a standard desktop workstation, for a $900 \times 270$ grid, as shown in Figs.~\ref{fig:2005}--\ref{fig:2068}. Reducing the grid size by a factor of~4 reduces typical computation times to about two hours, but the computation naturally lends itself to parallelization and can be done much faster, if need be.


\acknowledgements 
The work presented in this Letter was supported by the National Aeronautics and Space Administration under grants issued from the Living with a Star Targeted Research \& Technology Program (NNH08CC02C to RJL and NNX08AU30G to SWM). 
{\em SOHO} is a mission of international cooperation between ESA and NASA. The National Center for Atmospheric Research is sponsored by the National Science Foundation.




\clearpage

\end{document}